\documentclass{article}

\usepackage{microtype}
\usepackage{graphicx}
\usepackage{subfigure}
\usepackage{booktabs} % for professional tables

\usepackage{hyperref}

\usepackage[accepted]{ml4astro2025}

\usepackage{amsmath}
\usepackage{amssymb}
\usepackage{mathtools}
\usepackage{amsthm}
\usepackage{longtable} 
\usepackage{pdflscape}
\usepackage{gensymb}
\usepackage[capitalize,noabbrev]{cleveref}
\usepackage{xcolor}   % already loaded by tikz, but explicit doesn't hurt
\usepackage{tikz}
\usepackage{dashrule}
\definecolor{mpltab:blue}{HTML}{1f77b4}
\definecolor{mpltab:orange}{HTML}{ff7f0e}
\definecolor{mpltab:green}{HTML}{2ca02c}
\definecolor{mpltab:red}{HTML}{d62728}
\definecolor{mpltab:purple}{HTML}{9467bd}
\newcommand{\node}[1]{\mathsf{#1}}
\newcommand{\set}[1]{\mathbf{#1}}

\newcommand{\graph}{\mathcal{G}}

\newcommand{\parents}{\text{Pa}_{\graph}}

\theoremstyle{plain}

\theoremstyle{definition}

\theoremstyle{remark}

\usepackage[textsize=tiny]{todonotes}

\mlforastrotitlerunning{Causal Galactic Archaeology}

\begin{document}

\twocolumn[
\mlforastrotitle{Causal Discovery of Latent Variables in Galactic Archaeology}



\mlforastrosetsymbol{equal}{*}

\begin{mlforastroauthorlist}
%\icmlauthor{Anonymous Authors}{}
\mlforastroauthor{Zehao Jin*}{nyuad,fudan}
\mlforastroauthor{Yuxi(Lucy) Lu*}{osu}
\mlforastroauthor{Yuan-Sen Ting}{osu}
\mlforastroauthor{Yujia Zheng}{cmu}
\mlforastroauthor{Tobias Buck}{heidelberg}
\end{mlforastroauthorlist}

\mlforastroaffiliation{nyuad}{Center for Astrophysics and Space Science (CASS), New York University Abu Dhabi, PO Box 129188, Abu Dhabi, UAE}
\mlforastroaffiliation{fudan}{Center for Astronomy and Astrophysics and Department of Physics, Fudan University, Shanghai 200438, People’s Republic of China}
\mlforastroaffiliation{osu}{Center for Cosmology and Astroparticle Physics (CCAPP), The Ohio State University, 191 W. Woodruff Ave., Columbus, OH 43210, USA}
\mlforastroaffiliation{cmu}{Carnegie Mellon University, Pittsburgh, PA, USA}
\mlforastroaffiliation{heidelberg}{Universit\"at Heidelberg, Interdisziplin\"ares Zentrum f\"ur Wissenschaftliches Rechnen, Im Neuenheimer Feld 205, Heidelberg, Germany}

\mlforastrocorrespondingauthor{Zehao Jin}{zj448@nyu.edu}

\mlforastrokeywords{Causal Discovery, Galactic Archaeology, Astronomy}

\vskip 0.3in
]



\printAffiliationsAndNotice{\mlforastroEqualContribution} % otherwise use the standard text.

\begin{abstract}
Galactic archaeology—the study of stellar migration histories—provides insights into galaxy formation and evolution. However, establishing causal relationships between observable stellar properties and their birth conditions remains challenging, as key properties like birth radius are not directly observable. We employ Rank-based Latent Causal Discovery (RLCD) to uncover the causal structure governing the chemodynamics of a simulated Milky Way galaxy. Using only five observable properties (metallicity, age, and orbital parameters), we recover in a purely data-driven manner a causal graph containing two latent nodes that correspond to real physical properties: the birth radius and guiding radius of stars. Our study demonstrates the potential of causal discovery models in astrophysics.
\end{abstract}

\section{Introduction}\label{sec:intro}

The advent of large astronomical surveys has propelled much of the current interest in applying deep learning to astronomy. However, the observational nature of astronomy—where controlled experiments are impossible—often requires us to develop a deep understanding of the mechanistic causal relations governing astrophysical systems, beyond merely identifying correlations within observed variables.

\looseness=-1
Much of the advancement in astrophysics has relied on human intervention through forward modeling to match observations, and the statistical validation of physical models to decipher causal structures between fundamental drivers and emergent observations. This traditional approach, while successful, is inherently limited by human intuition and the complexity of the systems under study.

This raises the question: would it be possible to discover the underlying causal structure through automated causal graphical inference? Recent advances in causal discovery \citep[e.g.,][]{spirtes2000causation,pearl2009causality,pearl2016}—  which infers causal relationships from purely observational data—suggest this may indeed be feasible. In principle, latent variable models and directed acyclic graphs provide the statistical framework to distinguish between competing causal structures, potentially revealing relationships that human-guided analysis might overlook.

Causal discovery through probabilistic graphical models has a long history \citep{Geiger:1994,spirtes1995causal,spirtes2000causation,Chickering:2002,shimizu2006,zhang2008completeness,Huang:2018,deleu2022daggflownet}, and recently, physical sciences have begun to recognize its potential. \citep{runge2019inferring,li2020causal,scholkopf2021toward, zhang2024causal, yao2024marrying}. These successes demonstrate that causal discovery can reveal non-obvious relationships in complex systems where traditional approaches may struggle.

While astrophysical studies have been relatively late to adopt these methods, efforts to directly infer causal structures from astronomical data are beginning to emerge, including a preliminary pilot study by \citet{Pasquato2023prima} and Bayesian analyses of causal structures underlying galaxy–supermassive black hole coevolution \citep{Jin2025unravel}. However, due to the complexity of these astronomical systems and the indirect nature of observations, many inferred causal graph structures remain tentative and difficult to connect directly to known physical mechanisms.

\looseness=-1
To bridge this gap between causal discovery and physical understanding, we focus our investigation on galactic archaeology—the study of chemodynamic histories that shape the galaxies today. This field offers an ideal testbed for causal discovery methods because while the underlying physics is relatively well-understood, properties that govern stellar evolution—such as birth radius and historical orbital parameters—are not directly observable. These ``missing'' latent variables create a suitable scenario for causal discovery: we need methods that can both identify these hidden factors and establish their causal relationships with observable quantities.

\section{Methodology} \label{sec:methods}

To test whether causal discovery can recover physically meaningful latent variables in galactic archaeology, we require data where both observable properties and ground-truth birth conditions are available. Simulated galaxies provide this unique opportunity, allowing us to validate the causal structures discovered by our method against known physical quantities.

\subsection{Simulation Data}\label{sec:Simdata}
The simulated galaxy studied in this work is taken from the NIHAO-UHD project \citep{Buck2018, Buck2020b}, a set of high-resolution cosmological zoomed-in hydrodynamical simulations of Milky Way-mass galaxies. NIHAO (Numerical Investigation of a Hundred Astronomical Objects) comprises 100 simulated galaxies with halo masses ranging from $\sim 10^9-10^{12} M_\odot$. The dark matter halos were selected from a large-scale dark matter simulation based on an isolation criterion \citep[no similar mass companion within three virial radii at redshift $z=0$,][]{Wang2015}.

The NIHAO-UHD simulations employ a modified version of the smoothed particle hydrodynamics (SPH) solver GASOLINE2 \citep{Wadsley2017}, with star formation and feedback modeled following \citep{Stinson2006,Stinson2013}. These galaxies show agreement with observed Milky Way properties and local disk galaxies \citep{Obreja2019, Buck2020}, making them suitable analogs for our analysis. The simulations have been extensively validated in studies of MW-mass galaxies \citep[e.g.,][]{Buck2018, Buck2019b, Hilmi2020, Sestito2021, Obreja2022, Lu2022_rblim, Lu2022_turning, Wang2023}.

We focus on the g2.79e12 simulation. We select disk stars with [Fe/H] $>$ -1 located between 7-10 kpc at present day, a range chosen to ensure adequate sampling while matching typical observational surveys. To incorporate realistic observational uncertainties, we add uncertainties to age (10\%), [Fe/H] (0.02 dex), and [O/Fe] (0.06 dex).

Given the distinct formation histories of the high- and low-$\alpha$ disk populations in the Milky Way \citep{Bensby2014, Hayden2015}, we separate stars using the criterion [O/Fe] = -0.13[Fe/H] + 0.17. The high-$\alpha$ disk formed rapidly in a turbulent, gas-rich environment, while the low-$\alpha$ disk formed through gradual secular evolution \citep{Conroy2022, Xiang2022}. This pilot study focuses on the low-$\alpha$ disk, where secular processes dominate and causal relationships may be clearer to identify.

For our causal discovery analysis, we use five observable variables that trace stellar migration: metallicity [Fe/H], age, and three orbital parameters—vertical action $J_z$, angular momentum $L_z$, and eccentricity $e$. These quantities are routinely measured in spectroscopic surveys and encode information about both stellar birth conditions and subsequent dynamical evolution.  

\subsection{Discovery of Causal Structures}\label{sec:causal_discovery}

Having established our dataset, we now turn to the causal discovery analysis. Inferring causal relationships in astrophysical systems is complicated by unobserved (latent) variables—quantities that influence our observations but cannot be directly measured. These latent variables can confound relationships between observables or mediate their interactions, making causal inference challenging.

To address this challenge, we employ the Rank-based Latent Causal Discovery (RLCD) algorithm \citep{dong2024versatile}, which can uncover causal structures involving latent factors within linear systems. The key insight behind RLCD is that latent variables leave statistical signatures in the relationships between observed variables—specifically, they create rank deficiencies in the covariance matrix that can be detected and interpreted.

In our framework, we model the causal relationships using a Structural Causal Model (SCM). Consider a simple example: if birth radius (latent) influences both current metallicity and angular momentum (observed), then metallicity and angular momentum will be correlated through their shared latent cause. The SCM captures such relationships mathematically through a directed acyclic graph $\graph:=(\set{V}_\graph,\set{E}_\graph)$, where arrows represent causal influences. Each variable $V_i$ in this graph is generated according to a linear equation:
\begin{equation}
    \label{eq:lem}
\node{V}_i=\sum \nolimits_{\node{V}_j \in \parents(\node{V}_i)} a_{ij} \node{V}_j + \varepsilon_{\node{V}_i},
\end{equation}
where $\parents(V_i)$ denotes the variables that directly cause $V_i$ (its "parents" in the graph), $a_{ij}$ quantifies the strength of the causal effect from $V_j$ to $V_i$, and $\varepsilon_{\node{V}_i}$ represents random noise. The complete set of variables $\set{V}_\graph$ includes both observed variables $\set{X}_\graph$ (our five measured quantities) and latent variables $\set{L}_\graph$ (the hidden factors we aim to discover).

RLCD works by analyzing patterns in how observed variables covary. When latent variables are present, they constrain these covariation patterns in detectable ways. The algorithm identifies how many latent variables exist, which observed variables they influence, and whether latent variables influence each other. This enables reconstruction of a complete causal graph including both observed and hidden factors. The method has been validated on synthetic and real-world datasets \citep{dong2024versatile,dong2025parameter}.

\begin{figure}
\centering
\includegraphics[width=\linewidth]{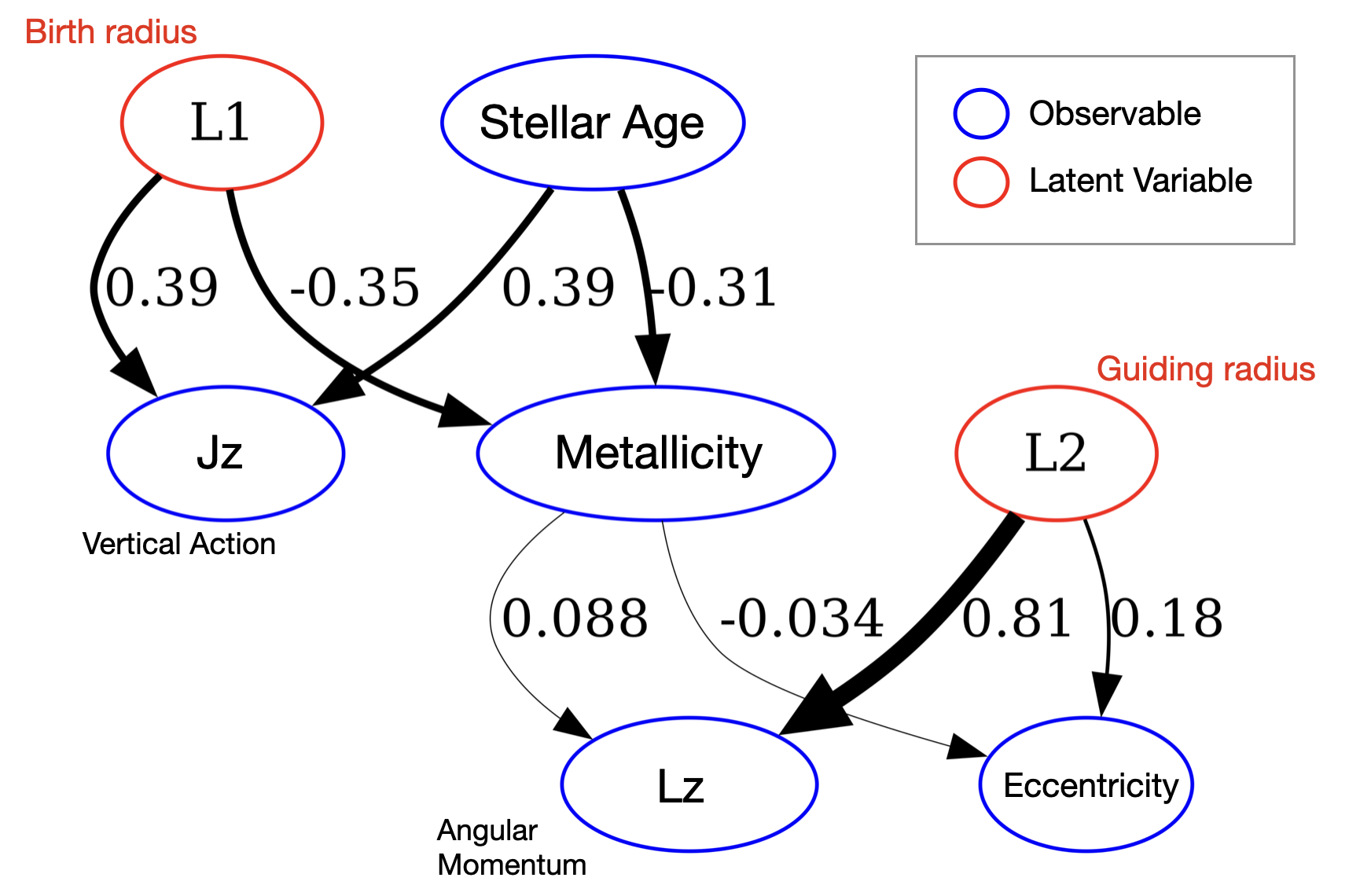}
\vspace{-0.5cm}
\caption{
Causal structure discovered by RLCD from five observable stellar properties. Blue ellipses represent observed variables, while red ellipses show the two latent variables identified by the algorithm. The analysis reveals that $L_1$ corresponds to birth radius and $L_2$ to guiding radius, as validated by comparison with ground truth. Numbers on arrows indicate the strength of causal relationships (edge coefficients $a_{ij}$) in the linear SCM.
}
\vspace{-0.3cm}
\label{fig:dag}
\end{figure}

After discovering the causal structure, we quantify the strength of each causal relationship by estimating the coefficients $a_{ij}$. As latent variables have no inherent scale, we fix the variance of each latent variable to unity—a standard convention that allows unique parameter estimation \citep{dong2025parameter}. We use maximum likelihood estimation to find the parameters that best explain our observed data given the discovered causal structure.

Finally, with the causal structure and parameters determined, we estimate the latent variable values for each individual star. This involves finding the latent values that best reconstruct the observed properties according to our linear model, minimizing the prediction error. 
%When the causal graph contains multiple interconnected latent variables, we estimate them hierarchically—first determining ``parent'' latent variables, then using these to estimate their ``children.'' 

\vspace{-0.3em}
\section{Results} \label{sec:results}

Applying RLCD to our five observable stellar properties yields the causal structure shown in Figure~\ref{fig:dag}. The algorithm identifies two latent variables, $L_1$ and $L_2$, whose connections to the observed variables provide insights into the underlying physics of stellar migration \citep{Sellwood2002}.

\begin{figure}
\centering
\includegraphics[width=\linewidth]{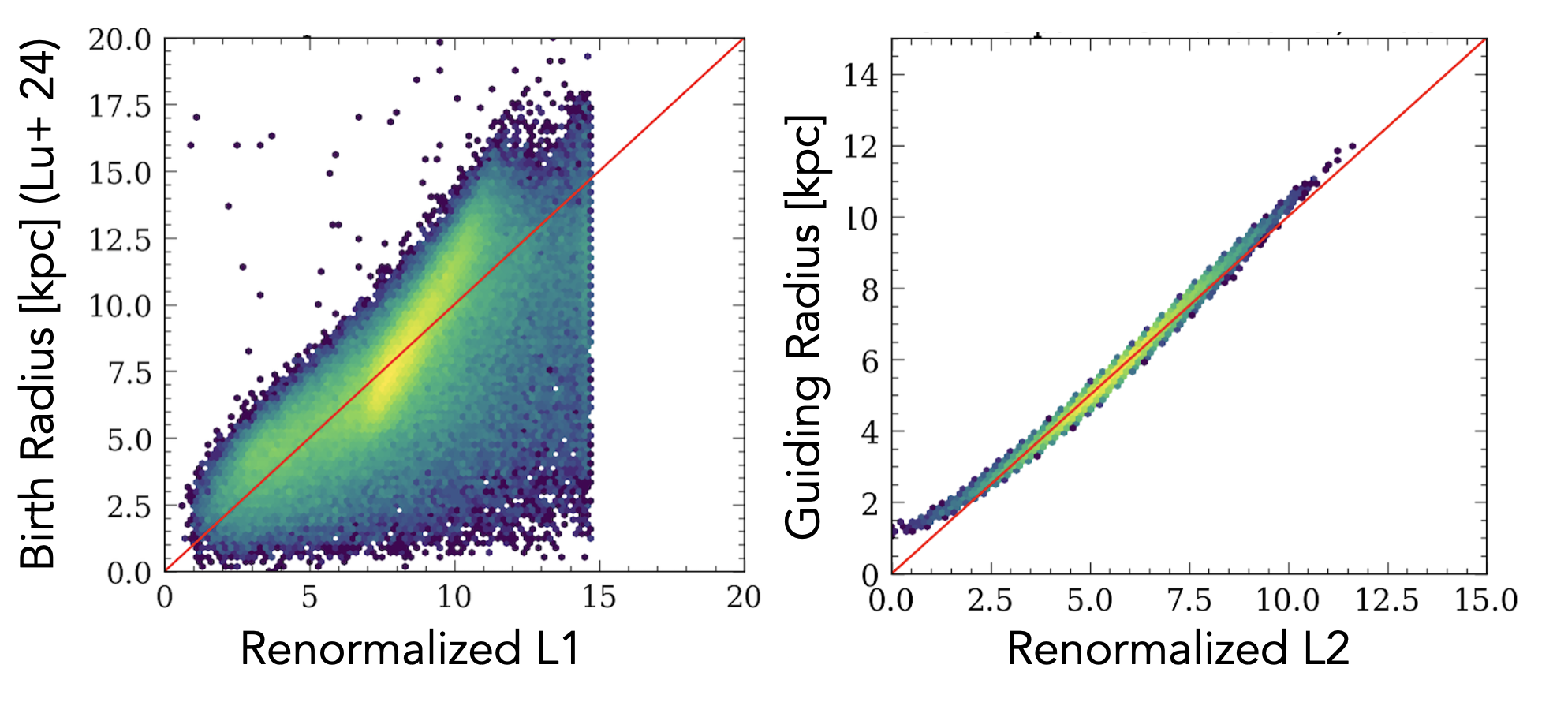}
\vspace{-0.8cm}
\caption{
Validation of discovered latent variables. \emph{Left}: Comparison between renormalized $L_1$ and birth radius inferred using the \citet{Lu2024} method. \emph{Right}: True guiding radius from simulation versus renormalized $L_2$. The discovered $L_1$ achieves comparable performance to \citep{Lu2024} for birth radius inference, while $L_2$ directly recovers the true guiding radius.
}
\vspace{-0.5cm}
\label{fig:L1L2}
\end{figure}

\begin{figure*}[ht!]
\centering
\includegraphics[width=0.75\linewidth]{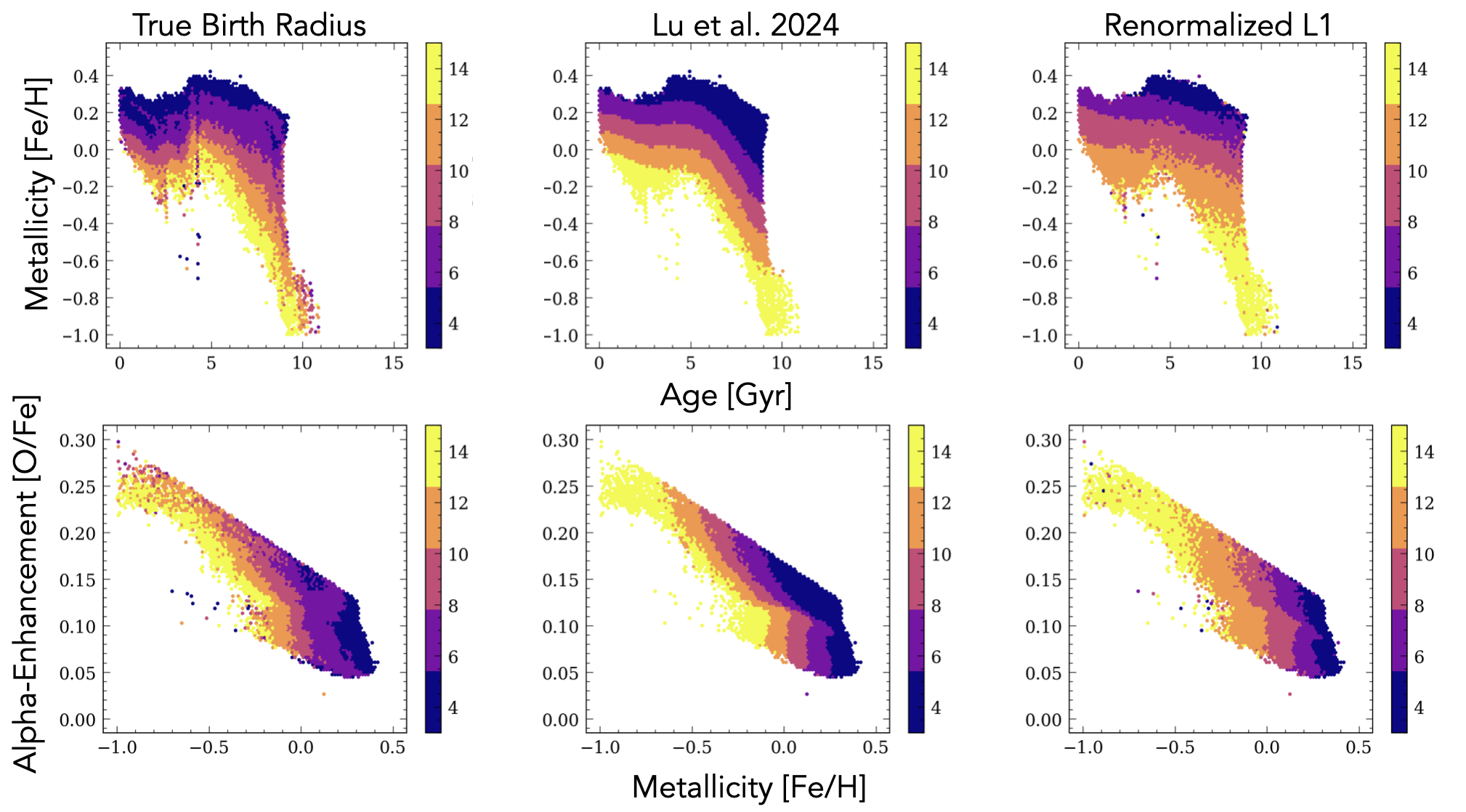}
\vspace{-0.3cm}
\caption{
Distribution of birth radii in chemical abundance space. \emph{Left}: True birth radius from simulation. \emph{Middle}: Birth radius inferred using the \citet{Lu2024} method. \emph{Right}: Renormalized $L_1$ from causal discovery. Top row shows the age-metallicity plane, bottom row shows the alpha-abundance plane. All three methods exhibit consistent patterns: younger, metal-rich stars originate from smaller galactic radii (yellow) while older, metal-poor stars come from larger radii (blue), reflecting inside-out galaxy formation. The similarity across columns validates that the discovered latent variable $L_1$ captures birth radius information despite using no prior knowledge.
}
\vspace{-0.3cm}
\label{fig:L1chem}
\end{figure*}

The discovered causal structure aligns well with our understanding of galactic chemical evolution. The latent variable $L_1$ influences both metallicity ([Fe/H]) and vertical action ($J_z$), while age shows similar causal connections. This pattern suggests that $L_1$ encodes information about stellar birth conditions—stars born at different galactic radii have distinct metallicities due to radial abundance gradients, and their vertical motions retain memory of their birth environments through the differential gravitational potential. Meanwhile, $L_2$ directly influences angular momentum ($L_z$) and eccentricity ($e$), the two quantities that together define a star's guiding radius. Stars can develop eccentric orbits through gravitational scattering with giant molecular clouds—a process known as ``blurring'' that preserves angular momentum while increasing eccentricity.

Based on these causal relationships, we hypothesize that $L_1$ corresponds to birth radius ($R_b$) and $L_2$ to guiding radius ($R_g$). Both quantities play fundamental roles in galactic dynamics: radial migration displaces stars from their birth radii over time \citep{Minchev2013, Sharma2021, Lu2024}, while orbital heating increases random motions around the guiding radius \citep{Sellwood2002}. Birth radius serves as the primary indicator of radial migration because it represents the initial condition from which stars have migrated, while guiding radius tracks the secular evolution of orbital eccentricity. Traditional methods estimate birth radii from age and metallicity under the assumption that the metallicity gradient is linear \citep{Minchev2018, Lu2024}, while guiding radii can be calculated directly from angular momentum. The causal structure discovered by RLCD independently recovers these known relationships.

To test our hypothesis, we leverage the simulation's complete stellar histories. After estimating individual stellar values for $L_1$ and $L_2$ using the identified parameters, we map these dimensionless quantities to physical units via polynomial transformations, accounting for the centering and standardization in our analysis. For $L_1$, we compare against birth radii inferred using the established method of \citet{Lu2024}, which estimates birth radii from [Fe/H] and age based on assumptions from \cite{Lu2022_rblim}. For $L_2$, we can directly compare against the true guiding radius from the simulation.

Figure~\ref{fig:L1L2} presents the validation results. For guiding radius (right panel), the agreement between $L_2$ and the ground truth is notable, with minimal scatter about the one-to-one relation. This tight correlation confirms that $L_2$ captures guiding radius—unsurprising given that guiding radius is directly encoded in the angular momentum that $L_2$ influences. For birth radius (left panel), we compare $L_1$ against the Lu et al. method. This consistency between two independent approaches—one theory-driven, one data-driven—suggests both methods capture similar physical information about stellar birth conditions.

Figure~\ref{fig:L1chem} provides further validation by examining birth radius distributions across chemical abundance space. All three approaches—ground truth, Lu et al. inference, and discovered $L_1$—show consistent patterns. In the age-[Fe/H] plane (top row), younger, metal-rich stars consistently appear at smaller birth radii, while older, metal-poor stars occupy larger radii, reflecting the inside-out growth of the galactic disk. The [O/Fe]-[Fe/H] plane (bottom row) reveals how radial abundance gradients at different epochs imprint on stellar chemistry. The discovered $L_1$ reproduces these chemical evolution patterns without incorporating any prior knowledge, demonstrating that RLCD extracts physically meaningful information from the data.

These results demonstrate that causal discovery can recover physically meaningful latent variables from observational data alone. The method not only identifies the correct number of hidden factors but also determines their proper causal relationships with observables. For guiding radius, the near-perfect recovery validates our approach. For birth radius, the agreement with established methods and consistency across chemical spaces confirms that causal discovery captures the same underlying physics that astronomers have uncovered through  theoretical work. While we have assumed linear relationships for this proof-of-concept, future work with nonlinear causal discovery methods may further improve the reconstruction of stellar birth properties.

\section{Broader Impact} \label{sec:conclusions}

\looseness=-1
This work demonstrates that causal discovery can uncover physically meaningful relationships in astrophysical systems from observational data alone. By successfully identifying birth radius and guiding radius as the latent variables governing stellar migration—without incorporating prior astrophysical knowledge—we validate the potential of automated causal inference in astronomy. The discovered structure not only aligns with decades of theoretical understanding but achieves predictive performance comparable to traditional domain-specific methods. This suggests that causal discovery can complement human-guided theoretical approaches, potentially revealing relationships that intuition might overlook in increasingly complex astronomical datasets.

\looseness=-1
Looking forward, our results establish a foundation for applying causal discovery across astronomy, where many phenomena involve unobservable latent variables—from dark matter shaping galaxy dynamics to stellar interiors driving evolution. As astronomical surveys grow in scale and dimensionality, such data-driven methods become essential for extracting physical insights. Future work extending to broader stellar populations and real observations will face additional challenges from selection effects and measurement uncertainties. The combination of causal discovery with flexible machine learning architectures promises interpretable AI systems that both predict and explain, guided by discovered causal structures rather than opaque correlations—moving toward a new paradigm where data-driven methods and physical understanding mutually reinforce our exploration of the physical processes that engender the universe.

\bibliography{bibliography}{}

\begin{thebibliography}{43}
\providecommand{\natexlab}[1]{#1}
\providecommand{\url}[1]{\texttt{#1}}
\expandafter\ifx\csname urlstyle\endcsname\relax
  \providecommand{\doi}[1]{doi: #1}\else
  \providecommand{\doi}{doi: \begingroup \urlstyle{rm}\Url}\fi

\bibitem[{Bensby} et~al.(2014){Bensby}, {Feltzing}, and {Oey}]{Bensby2014}
{Bensby}, T., {Feltzing}, S., and {Oey}, M.~S.
\newblock {Exploring the Milky Way stellar disk. A detailed elemental abundance study of 714 F and G dwarf stars in the solar neighbourhood}.
\newblock \emph{\aap}, 562:\penalty0 A71, February 2014.
\newblock \doi{10.1051/0004-6361/201322631}.

\bibitem[Buck(2019)]{Buck2020}
Buck, T.
\newblock {On the origin of the chemical bimodality of disc stars: a tale of merger and migration}.
\newblock \emph{Monthly Notices of the Royal Astronomical Society}, 491\penalty0 (4):\penalty0 5435--5446, 11 2019.
\newblock ISSN 0035-8711.
\newblock \doi{10.1093/mnras/stz3289}.
\newblock URL \url{https://doi.org/10.1093/mnras/stz3289}.

\bibitem[{Buck} et~al.(2018){Buck}, {Ness}, {Macci{\`o}}, {Obreja}, and {Dutton}]{Buck2018}
{Buck}, T., {Ness}, M.~K., {Macci{\`o}}, A.~V., {Obreja}, A., and {Dutton}, A.~A.
\newblock {Stars Behind Bars. I. The Milky Way's Central Stellar Populations}.
\newblock \emph{\apj}, 861\penalty0 (2):\penalty0 88, July 2018.
\newblock \doi{10.3847/1538-4357/aac890}.

\bibitem[{Buck} et~al.(2019){Buck}, {Ness}, {Obreja}, {Macci{\`o}}, and {Dutton}]{Buck2019b}
{Buck}, T., {Ness}, M., {Obreja}, A., {Macci{\`o}}, A.~V., and {Dutton}, A.~A.
\newblock {Stars behind Bars II: A Cosmological Formation Scenario for the Milky Way{\textquoteright}s Central Stellar Structure}.
\newblock \emph{\apj}, 874\penalty0 (1):\penalty0 67, March 2019.
\newblock \doi{10.3847/1538-4357/aaffd0}.

\bibitem[{Buck} et~al.(2020){Buck}, {Obreja}, {Macci{\`o}}, {Minchev}, {Dutton}, and {Ostriker}]{Buck2020b}
{Buck}, T., {Obreja}, A., {Macci{\`o}}, A.~V., {Minchev}, I., {Dutton}, A.~A., and {Ostriker}, J.~P.
\newblock {NIHAO-UHD: the properties of MW-like stellar discs in high-resolution cosmological simulations}.
\newblock \emph{\mnras}, 491\penalty0 (3):\penalty0 3461--3478, January 2020.
\newblock \doi{10.1093/mnras/stz3241}.

\bibitem[Chickering(2002)]{Chickering:2002}
Chickering, D.
\newblock Optimal structure identification with greedy search.
\newblock \emph{Journal of Machine Learning Research}, 3:\penalty0 507--554, 01 2002.
\newblock \doi{10.1162/153244303321897717}.

\bibitem[{Conroy} et~al.(2022){Conroy}, {Weinberg}, {Naidu}, {Buck}, {Johnson}, {Cargile}, {Bonaca}, {Caldwell}, {Chandra}, {Han}, {Johnson}, {Speagle}, {Ting}, {Woody}, and {Zaritsky}]{Conroy2022}
{Conroy}, C., {Weinberg}, D.~H., {Naidu}, R.~P., {Buck}, T., {Johnson}, J.~W., {Cargile}, P., {Bonaca}, A., {Caldwell}, N., {Chandra}, V., {Han}, J.~J., {Johnson}, B.~D., {Speagle}, J.~S., {Ting}, Y.-S., {Woody}, T., and {Zaritsky}, D.
\newblock {Birth of the Galactic Disk Revealed by the H3 Survey}.
\newblock \emph{arXiv e-prints}, art. arXiv:2204.02989, April 2022.
\newblock \doi{10.48550/arXiv.2204.02989}.

\bibitem[Deleu et~al.(2022)Deleu, G{\'o}is, Emezue, Rankawat, Lacoste-Julien, Bauer, and Bengio]{deleu2022daggflownet}
Deleu, T., G{\'o}is, A., Emezue, C.~C., Rankawat, M., Lacoste-Julien, S., Bauer, S., and Bengio, Y.
\newblock Bayesian structure learning with generative flow networks.
\newblock In \emph{The 38th Conference on Uncertainty in Artificial Intelligence}, 2022.
\newblock URL \url{https://openreview.net/forum?id=HElfed8j9g9}.

\bibitem[Dong et~al.(2024)Dong, Huang, Ng, Song, Zheng, Jin, Legaspi, Spirtes, and Zhang]{dong2024versatile}
Dong, X., Huang, B., Ng, I., Song, X., Zheng, Y., Jin, S., Legaspi, R., Spirtes, P., and Zhang, K.
\newblock A versatile causal discovery framework to allow causally-related hidden variables.
\newblock In \emph{The Twelfth International Conference on Learning Representations}, 2024.

\bibitem[Dong et~al.(2025)Dong, Ng, Huang, Sun, Jin, Legaspi, Spirtes, and Zhang]{dong2025parameter}
Dong, X., Ng, I., Huang, B., Sun, Y., Jin, S., Legaspi, R., Spirtes, P., and Zhang, K.
\newblock On the parameter identifiability of partially observed linear causal models.
\newblock \emph{Advances in Neural Information Processing Systems}, 37:\penalty0 30740--30771, 2025.

\bibitem[Geiger \& Heckerman(1994)Geiger and Heckerman]{Geiger:1994}
Geiger, D. and Heckerman, D.
\newblock Learning gaussian networks.
\newblock In \emph{Proceedings of the Tenth International Conference on Uncertainty in Artificial Intelligence}, UAI'94, pp.\  235--243, San Francisco, CA, USA, 1994. Morgan Kaufmann Publishers Inc.
\newblock ISBN 1558603328.

\bibitem[{Hayden} et~al.(2015){Hayden}, {Bovy}, {Holtzman}, {Nidever}, {Bird}, {Weinberg}, {Andrews}, {Majewski}, {Allende Prieto}, {Anders}, {Beers}, {Bizyaev}, {Chiappini}, {Cunha}, {Frinchaboy}, {Garc{\'\i}a-Her{\'n}andez}, {Garc{\'\i}a P{\'e}rez}, {Girardi}, {Harding}, {Hearty}, {Johnson}, {M{\'e}sz{\'a}ros}, {Minchev}, {O'Connell}, {Pan}, {Robin}, {Schiavon}, {Schneider}, {Schultheis}, {Shetrone}, {Skrutskie}, {Steinmetz}, {Smith}, {Wilson}, {Zamora}, and {Zasowski}]{Hayden2015}
{Hayden}, M.~R., {Bovy}, J., {Holtzman}, J.~A., {Nidever}, D.~L., {Bird}, J.~C., {Weinberg}, D.~H., {Andrews}, B.~H., {Majewski}, S.~R., {Allende Prieto}, C., {Anders}, F., {Beers}, T.~C., {Bizyaev}, D., {Chiappini}, C., {Cunha}, K., {Frinchaboy}, P., {Garc{\'\i}a-Her{\'n}andez}, D.~A., {Garc{\'\i}a P{\'e}rez}, A.~E., {Girardi}, L., {Harding}, P., {Hearty}, F.~R., {Johnson}, J.~A., {M{\'e}sz{\'a}ros}, S., {Minchev}, I., {O'Connell}, R., {Pan}, K., {Robin}, A.~C., {Schiavon}, R.~P., {Schneider}, D.~P., {Schultheis}, M., {Shetrone}, M., {Skrutskie}, M., {Steinmetz}, M., {Smith}, V., {Wilson}, J.~C., {Zamora}, O., and {Zasowski}, G.
\newblock {Chemical Cartography with APOGEE: Metallicity Distribution Functions and the Chemical Structure of the Milky Way Disk}.
\newblock \emph{\apj}, 808\penalty0 (2):\penalty0 132, August 2015.
\newblock \doi{10.1088/0004-637X/808/2/132}.

\bibitem[{Hilmi} et~al.(2020){Hilmi}, {Minchev}, {Buck}, {Martig}, {Quillen}, {Monari}, {Famaey}, {de Jong}, {Laporte}, {Read}, {Sanders}, {Steinmetz}, and {Wegg}]{Hilmi2020}
{Hilmi}, T., {Minchev}, I., {Buck}, T., {Martig}, M., {Quillen}, A.~C., {Monari}, G., {Famaey}, B., {de Jong}, R.~S., {Laporte}, C.~F.~P., {Read}, J., {Sanders}, J.~L., {Steinmetz}, M., and {Wegg}, C.
\newblock {Fluctuations in galactic bar parameters due to bar-spiral interaction}.
\newblock \emph{\mnras}, 497\penalty0 (1):\penalty0 933--955, September 2020.
\newblock \doi{10.1093/mnras/staa1934}.

\bibitem[Huang et~al.(2018)Huang, Zhang, Lin, Sch\"{o}lkopf, and Glymour]{Huang:2018}
Huang, B., Zhang, K., Lin, Y., Sch\"{o}lkopf, B., and Glymour, C.
\newblock Generalized score functions for causal discovery.
\newblock In \emph{Proceedings of the 24th ACM SIGKDD International Conference on Knowledge Discovery \& Data Mining}, KDD '18, pp.\  1551--1560, New York, NY, USA, 2018. Association for Computing Machinery.
\newblock ISBN 9781450355520.
\newblock \doi{10.1145/3219819.3220104}.
\newblock URL \url{https://doi.org/10.1145/3219819.3220104}.

\bibitem[Jin et~al.(2025)Jin, Pasquato, Davis, Deleu, Luo, Cho, Lemos, Perreault-Levasseur, Bengio, Kang, Macciò, and Hezaveh]{Jin2025unravel}
Jin, Z., Pasquato, M., Davis, B.~L., Deleu, T., Luo, Y., Cho, C., Lemos, P., Perreault-Levasseur, L., Bengio, Y., Kang, X., Macciò, A.~V., and Hezaveh, Y.
\newblock Causal discovery in astrophysics: Unraveling supermassive black hole and galaxy coevolution.
\newblock \emph{The Astrophysical Journal}, 979\penalty0 (2):\penalty0 212, jan 2025.
\newblock \doi{10.3847/1538-4357/ad9ded}.
\newblock URL \url{https://dx.doi.org/10.3847/1538-4357/ad9ded}.

\bibitem[Li et~al.(2020)Li, Torralba, Anandkumar, Fox, and Garg]{li2020causal}
Li, Y., Torralba, A., Anandkumar, A., Fox, D., and Garg, A.
\newblock Causal discovery in physical systems from videos.
\newblock \emph{Advances in Neural Information Processing Systems}, 33:\penalty0 9180--9192, 2020.

\bibitem[{Lu} et~al.(2022{\natexlab{a}}){Lu}, {Buck}, {Minchev}, and {Ness}]{Lu2022_rblim}
{Lu}, Y., {Buck}, T., {Minchev}, I., and {Ness}, M.~K.
\newblock {Reliability and limitations of inferring birth radii in the Milky Way disc}.
\newblock \emph{\mnras}, 515\penalty0 (1):\penalty0 L34--L38, September 2022{\natexlab{a}}.
\newblock \doi{10.1093/mnrasl/slac065}.

\bibitem[{Lu} et~al.(2022{\natexlab{b}}){Lu}, {Ness}, {Buck}, and {Carr}]{Lu2022_turning}
{Lu}, Y.~L., {Ness}, M.~K., {Buck}, T., and {Carr}, C.
\newblock {Turning points in the age-metallicity relations - created by late satellite infall and enhanced by radial migration}.
\newblock \emph{\mnras}, 512\penalty0 (4):\penalty0 4697--4714, June 2022{\natexlab{b}}.
\newblock \doi{10.1093/mnras/stac780}.

\bibitem[{Lu} et~al.(2024){Lu}, {Minchev}, {Buck}, {Khoperskov}, {Steinmetz}, {Libeskind}, {Cescutti}, {Freeman}, and {Ratcliffe}]{Lu2024}
{Lu}, Y.~L., {Minchev}, I., {Buck}, T., {Khoperskov}, S., {Steinmetz}, M., {Libeskind}, N., {Cescutti}, G., {Freeman}, K.~C., and {Ratcliffe}, B.
\newblock {There is no place like home - finding birth radii of stars in the Milky Way}.
\newblock \emph{\mnras}, 535\penalty0 (1):\penalty0 392--405, November 2024.
\newblock \doi{10.1093/mnras/stae2364}.

\bibitem[{Minchev} et~al.(2013){Minchev}, {Chiappini}, and {Martig}]{Minchev2013}
{Minchev}, I., {Chiappini}, C., and {Martig}, M.
\newblock {Chemodynamical evolution of the Milky Way disk. I. The solar vicinity}.
\newblock \emph{\aap}, 558:\penalty0 A9, October 2013.
\newblock \doi{10.1051/0004-6361/201220189}.

\bibitem[{Minchev} et~al.(2018){Minchev}, {Anders}, {Recio-Blanco}, {Chiappini}, {de Laverny}, {Queiroz}, {Steinmetz}, {Adibekyan}, {Carrillo}, {Cescutti}, {Guiglion}, {Hayden}, {de Jong}, {Kordopatis}, {Majewski}, {Martig}, and {Santiago}]{Minchev2018}
{Minchev}, I., {Anders}, F., {Recio-Blanco}, A., {Chiappini}, C., {de Laverny}, P., {Queiroz}, A., {Steinmetz}, M., {Adibekyan}, V., {Carrillo}, I., {Cescutti}, G., {Guiglion}, G., {Hayden}, M., {de Jong}, R.~S., {Kordopatis}, G., {Majewski}, S.~R., {Martig}, M., and {Santiago}, B.~X.
\newblock {Estimating stellar birth radii and the time evolution of Milky Way's ISM metallicity gradient}.
\newblock \emph{\mnras}, 481\penalty0 (2):\penalty0 1645--1657, December 2018.
\newblock \doi{10.1093/mnras/sty2033}.

\bibitem[{Obreja} et~al.(2019){Obreja}, {Dutton}, {Macci{\`o}}, {Moster}, {Buck}, {van de Ven}, {Wang}, {Stinson}, and {Zhu}]{Obreja2019}
{Obreja}, A., {Dutton}, A.~A., {Macci{\`o}}, A.~V., {Moster}, B., {Buck}, T., {van de Ven}, G., {Wang}, L., {Stinson}, G.~S., and {Zhu}, L.
\newblock {NIHAO XVI: the properties and evolution of kinematically selected discs, bulges, and stellar haloes}.
\newblock \emph{\mnras}, 487\penalty0 (3):\penalty0 4424--4456, August 2019.
\newblock \doi{10.1093/mnras/stz1563}.

\bibitem[{Obreja} et~al.(2022){Obreja}, {Buck}, and {Macci{\`o}}]{Obreja2022}
{Obreja}, A., {Buck}, T., and {Macci{\`o}}, A.~V.
\newblock {A first estimate of the Milky Way dark matter halo spin}.
\newblock \emph{\aap}, 657:\penalty0 A15, January 2022.
\newblock \doi{10.1051/0004-6361/202140983}.

\bibitem[{Pasquato} et~al.(2023){Pasquato}, {Jin}, {Lemos}, {Davis}, and {Macci{\`o}}]{Pasquato2023prima}
{Pasquato}, M., {Jin}, Z., {Lemos}, P., {Davis}, B.~L., and {Macci{\`o}}, A.~V.
\newblock {Causa prima: cosmology meets causal discovery for the first time}.
\newblock \emph{arXiv e-prints}, art. arXiv:2311.15160, November 2023.
\newblock \doi{10.48550/arXiv.2311.15160}.

\bibitem[Pearl(2009)]{pearl2009causality}
Pearl, J.
\newblock \emph{Causality}.
\newblock Cambridge university press, 2009.

\bibitem[Pearl et~al.(2016)Pearl, Glymour, and Jewell]{pearl2016}
Pearl, J., Glymour, M., and Jewell, N.~P.
\newblock \emph{Causal inference in statistics : a primer}.
\newblock Wiley, Chichester, West Sussex, 2016.
\newblock ISBN 9781119186854.

\bibitem[Runge et~al.(2019)Runge, Bathiany, Bollt, Camps-Valls, Coumou, Deyle, Glymour, Kretschmer, Mahecha, Mu{\~n}oz-Mar{\'\i}, et~al.]{runge2019inferring}
Runge, J., Bathiany, S., Bollt, E., Camps-Valls, G., Coumou, D., Deyle, E., Glymour, C., Kretschmer, M., Mahecha, M.~D., Mu{\~n}oz-Mar{\'\i}, J., et~al.
\newblock Inferring causation from time series in earth system sciences.
\newblock \emph{Nature communications}, 10\penalty0 (1):\penalty0 2553, 2019.

\bibitem[Sch{\"o}lkopf et~al.(2021)Sch{\"o}lkopf, Locatello, Bauer, Ke, Kalchbrenner, Goyal, and Bengio]{scholkopf2021toward}
Sch{\"o}lkopf, B., Locatello, F., Bauer, S., Ke, N.~R., Kalchbrenner, N., Goyal, A., and Bengio, Y.
\newblock Toward causal representation learning.
\newblock \emph{Proceedings of the IEEE}, 109\penalty0 (5):\penalty0 612--634, 2021.

\bibitem[{Sellwood} \& {Binney}(2002){Sellwood} and {Binney}]{Sellwood2002}
{Sellwood}, J.~A. and {Binney}, J.~J.
\newblock {Radial mixing in galactic discs}.
\newblock \emph{\mnras}, 336\penalty0 (3):\penalty0 785--796, November 2002.
\newblock \doi{10.1046/j.1365-8711.2002.05806.x}.

\bibitem[{Sestito} et~al.(2021){Sestito}, {Buck}, {Starkenburg}, {Martin}, {Navarro}, {Venn}, {Obreja}, {Jablonka}, and {Macci{\`o}}]{Sestito2021}
{Sestito}, F., {Buck}, T., {Starkenburg}, E., {Martin}, N.~F., {Navarro}, J.~F., {Venn}, K.~A., {Obreja}, A., {Jablonka}, P., and {Macci{\`o}}, A.~V.
\newblock {Exploring the origin of low-metallicity stars in Milky-Way-like galaxies with the NIHAO-UHD simulations}.
\newblock \emph{\mnras}, 500\penalty0 (3):\penalty0 3750--3762, January 2021.
\newblock \doi{10.1093/mnras/staa3479}.

\bibitem[{Sharma} et~al.(2021){Sharma}, {Hayden}, and {Bland-Hawthorn}]{Sharma2021}
{Sharma}, S., {Hayden}, M.~R., and {Bland-Hawthorn}, J.
\newblock {Chemical enrichment and radial migration in the Galactic disc - the origin of the [{\ensuremath{\alpha}}Fe] double sequence}.
\newblock \emph{\mnras}, 507\penalty0 (4):\penalty0 5882--5901, November 2021.
\newblock \doi{10.1093/mnras/stab2015}.

\bibitem[Shimizu et~al.(2006)Shimizu, Hoyer, Hyvrinen, and Kerminen]{shimizu2006}
Shimizu, S., Hoyer, P.~O., Hyvrinen, A., and Kerminen, A.~J.
\newblock A linear non-gaussian acyclic model for causal discovery.
\newblock \emph{Journal of Machine Learning Research}, 7\penalty0 (4):\penalty0 2003--2030, 2006.

\bibitem[Spirtes et~al.(1995)Spirtes, Meek, and Richardson]{spirtes1995causal}
Spirtes, P., Meek, C., and Richardson, T.
\newblock Causal inference in the presence of latent variables and selection bias.
\newblock In \emph{Proceedings of the Eleventh Conference on Uncertainty in Artificial Intelligence}, pp.\  499--506, August 1995.

\bibitem[Spirtes et~al.(2001)Spirtes, Glymour, and Scheines]{spirtes2000causation}
Spirtes, P., Glymour, C., and Scheines, R.
\newblock \emph{Causation, Prediction, and Search}.
\newblock The MIT Press, 01 2001.
\newblock ISBN 9780262284158.
\newblock \doi{10.7551/mitpress/1754.001.0001}.
\newblock URL \url{https://doi.org/10.7551/mitpress/1754.001.0001}.

\bibitem[{Stinson} et~al.(2006){Stinson}, {Seth}, {Katz}, {Wadsley}, {Governato}, and {Quinn}]{Stinson2006}
{Stinson}, G., {Seth}, A., {Katz}, N., {Wadsley}, J., {Governato}, F., and {Quinn}, T.
\newblock {Star formation and feedback in smoothed particle hydrodynamic simulations - I. Isolated galaxies}.
\newblock \emph{\mnras}, 373\penalty0 (3):\penalty0 1074--1090, December 2006.
\newblock \doi{10.1111/j.1365-2966.2006.11097.x}.

\bibitem[{Stinson} et~al.(2013){Stinson}, {Bovy}, {Rix}, {Brook}, {Ro{\v{s}}kar}, {Dalcanton}, {Macci{\`o}}, {Wadsley}, {Couchman}, and {Quinn}]{Stinson2013}
{Stinson}, G.~S., {Bovy}, J., {Rix}, H.~W., {Brook}, C., {Ro{\v{s}}kar}, R., {Dalcanton}, J.~J., {Macci{\`o}}, A.~V., {Wadsley}, J., {Couchman}, H.~M.~P., and {Quinn}, T.~R.
\newblock {MaGICC thick disc - I. Comparing a simulated disc formed with stellar feedback to the Milky Way}.
\newblock \emph{\mnras}, 436\penalty0 (1):\penalty0 625--634, November 2013.
\newblock \doi{10.1093/mnras/stt1600}.

\bibitem[{Wadsley} et~al.(2017){Wadsley}, {Keller}, and {Quinn}]{Wadsley2017}
{Wadsley}, J.~W., {Keller}, B.~W., and {Quinn}, T.~R.
\newblock {Gasoline2: a modern smoothed particle hydrodynamics code}.
\newblock \emph{\mnras}, 471\penalty0 (2):\penalty0 2357--2369, October 2017.
\newblock \doi{10.1093/mnras/stx1643}.

\bibitem[{Wang} et~al.(2023){Wang}, {Carrillo}, {Ness}, and {Buck}]{Wang2023}
{Wang}, K., {Carrillo}, A., {Ness}, M.~K., and {Buck}, T.
\newblock {The individual abundance distributions of disc stars across birth radii in GALAH}.
\newblock \emph{arXiv e-prints}, art. arXiv:2307.04724, July 2023.
\newblock \doi{10.48550/arXiv.2307.04724}.

\bibitem[{Wang} et~al.(2015){Wang}, {Dutton}, {Stinson}, {Macci{\`o}}, {Penzo}, {Kang}, {Keller}, and {Wadsley}]{Wang2015}
{Wang}, L., {Dutton}, A.~A., {Stinson}, G.~S., {Macci{\`o}}, A.~V., {Penzo}, C., {Kang}, X., {Keller}, B.~W., and {Wadsley}, J.
\newblock {NIHAO project - I. Reproducing the inefficiency of galaxy formation across cosmic time with a large sample of cosmological hydrodynamical simulations}.
\newblock \emph{\mnras}, 454\penalty0 (1):\penalty0 83--94, November 2015.
\newblock \doi{10.1093/mnras/stv1937}.

\bibitem[{Xiang} \& {Rix}(2022){Xiang} and {Rix}]{Xiang2022}
{Xiang}, M. and {Rix}, H.-W.
\newblock {A time-resolved picture of our Milky Way's early formation history}.
\newblock \emph{\nat}, 603\penalty0 (7902):\penalty0 599--603, March 2022.
\newblock \doi{10.1038/s41586-022-04496-5}.

\bibitem[Yao et~al.(2024)Yao, Muller, and Locatello]{yao2024marrying}
Yao, D., Muller, C., and Locatello, F.
\newblock Marrying causal representation learning with dynamical systems for science.
\newblock In \emph{The Thirty-eighth Annual Conference on Neural Information Processing Systems}, 2024.

\bibitem[Zhang(2008)]{zhang2008completeness}
Zhang, J.
\newblock On the completeness of orientation rules for causal discovery in the presence of latent confounders and selection bias.
\newblock \emph{Artificial Intelligence}, 172\penalty0 (16-17):\penalty0 1873--1896, 2008.

\bibitem[Zhang et~al.(2024)Zhang, Xie, Ng, and Zheng]{zhang2024causal}
Zhang, K., Xie, S., Ng, I., and Zheng, Y.
\newblock Causal representation learning from multiple distributions: A general setting.
\newblock In \emph{International Conference on Machine Learning}, pp.\  60057--60075. PMLR, 2024.

\end{thebibliography}
\bibliographystyle{icml2025}

%%%%%%%%%%%%%%%%%%%%%%%%%%%%%%%%%%%%%%%%%%%%%%%%%%%%%%%%%%%%%%%%%%%%%%%%%%%%%%%
%%%%%%%%%%%%%%%%%%%%%%%%%%%%%%%%%%%%%%%%%%%%%%%%%%%%%%%%%%%%%%%%%%%%%%%%%%%%%%%
% APPENDIX
%%%%%%%%%%%%%%%%%%%%%%%%%%%%%%%%%%%%%%%%%%%%%%%%%%%%%%%%%%%%%%%%%%%%%%%%%%%%%%%
%%%%%%%%%%%%%%%%%%%%%%%%%%%%%%%%%%%%%%%%%%%%%%%%%%%%%%%%%%%%%%%%%%%%%%%%%%%%%%%
\newpage
\appendix
\onecolumn

\end{document}